**Room temperature reversible colossal volto-magnetic effect in all-oxide metallic-magnet/topotactic-phase-transition material heterostructures**

Sourav Chowdhury,[1,2] Supriyo Majumder,[1,3] Rajan Mishra,[1] Arup Kumar Mandal,[1,4] Anita Bagri,[1] Satish Yadav,[1] Suman Karmarkar,[1] D. M. Phase,[1] & R. J. Choudhary[1*]

[1]UGC-DAE Consortium for Scientific Research, University Campus, Khandwa Road, Indore, Madhya Pradesh 452001, India.

[2]Deutsches Elektronen-Synchrotron DESY, Notkestraβe 85, 22607 Hamburg, Germany.

[3]Northwestern University, Cook Hall, 2220 Campus Drive, Evanston, IIIinois 60208 USA.

[4]Indian Institute of Science, CV Raman Rd, Bengaluru, Karnataka 560012, India.

Multiferroic materials have undergone extensive research in the past two decades in an effort to produce a sizable room-temperature magneto-electric (ME) effect in either exclusive or composite materials for use in a variety of electronic or spintronic devices. These studies have looked into the ME effect by switching the electric polarization by the magnetic field or switching the magnetism by the electric field. Here, an innovative way is developed to knot the functional properties based on the tremendous modulation of electronics and magnetization by the electric field of the topotactic phase transitions (TPT) in heterostructures composed of metallic-magnet/TPT-material. It is divulged that application of a nominal potential difference of 2-3 Volts induces gigantic changes in magnetization by 100-250% leading to colossal Volto-magnetic effect, which would be tremendously beneficial for low-power consumption applications in spintronics. Switching electronics and magnetism by inducing TPT through applying an electric field requires much less energy, making such TPT-based systems promising for energy-efficient memory and logic applications as well as opening a plethora of tremendous opportunities for applications in different domains.

*Corresponding Author

E-mail: ram@csr.res.in

Over the last couple of decades, there has been a surge of research expeditions on the multiferroic materials proclaiming the magneto-electric (ME) effect, i.e., the control of a ferroelectric property via magnetic field or control of a magnetic property via an electric field, owing to their great potential applications in sensors, actuators, novel spintronic devices, magnetic read/electric write hard disk[1-5]. However, such materials are rare due to mutually exclusive qualifications of the ferroelectric (FE) and ferromagnetic (FM) ordering. The single-phase ME materials categorized as type-I and type-II exhibit either a very weak ME effect at room temperature or exhibit somewhat strong ME effect albeit at low-temperature values below 100K, which limits their utility for a practical device application[6-8]. Hence the search is extended to design composite materials consisting of FE and FM materials or their heterostructures for higher ME effect, via coupling between the piezo/ferro-electric/magnetic interaction at the contact interface[9-12]. Four main mechanisms are reported in the literature as responsible for the interfacial ME coupling: charge accumulation or depletion at the interface[13-19], strain-mediated effects[20-24], ion migration[25], and morphological changes[26,27]. However, so far, a realistic ME effect in composite structures for device applications is yet to be realized. One of the major challenges is the dependence of the ME effect on the interfacial connected networks of FE and FM regions of the composite systems. Also, the leaking nature of the FE regions in the thin film heterostructures inhibits the proper coupling at the interface. Therefore, it is highly desired to look for an alternative architecture yielding a strong ME effect without worrying about the leakage in the FE layer at the heterointerface.

Here, we propose an innovative way to tune the functional properties based on the modulation by the electric field of topotactic phase transitions (TPT) in heterostructures composed of metallic-magnet/TPT-material, as shown schematically in Fig. 1(a). TPT is a reversible phase transition between sharply contrasted structural, electronic, magnetic, electrical, and optical phases of a material, which can be induced by applying a very small external electric field[28-30]. We envisioned that in such a heterostructure device, the TPT-material undergoes a completely reversible structural, electronic, magnetic, electrical, and optical phase transition owing to the Mott-switching under the application of a very small electric bias wherein the metallic-magnet remains unchanged. As the TPT-layer undergoes a complete phase transition, the coupled electronic and magnetic properties with the metallic-magnetic layer will also be drastically modulated via ME coupling. Thus, such metallic-magnet/TPT-material heterostructure system can be tested for the coupled ME (spintronics) applications.

In the present study, we demonstrate a colossal converse ME effect as high as 250% at room temperature at an application of a minuscule electric voltage, which causes dramatic changes in the magnetization values in a new class of all-oxides magnetic heterostructure system. We coin the term volto-magnetic (VM) effect in the studied all magnetic oxides heterostructure to make it distinct from the generally observed ME effect in FE-FM composite multiferroic systems owing to the different mechanisms involved in the phenomena. The presented study is a concept of proof on a $SrRuO_3/SrCoO_x$ heterostructure grown on conducting Nb-doped $SrTiO_3$ (STO) substrate, which can be extended to many such oxide heterostructures overcoming the bottleneck of the limited ME effect and its application landscape. The colossal VM effect arises due to the electric field tuneable magnetic ground state of $SrCoO_x$ ($SCO_x$) in conjunction with $SrRuO_3$ (SRO). The observed VM effect is exotic and unmatched by any previously reported ME effect in any multiferroic system.

The topotactic phase evolution in $SCO_x$ between the insulating-antiferromagnet x=2.5 and the metallic-ferromagnet x=3 end members promote these materials in a wide range of applications, such as solid oxide fuel cell, neuromorphic computation, Mottronics, memory devices, and smart window materials[30-37]. The entwined structural, electronic, and magnetic properties of $SrCoO_{2.5}$ ($SCO_{2.5}$) demonstrate a fascinating tunable electrical and magnetic phase via strain in thin film form. In the bulk or strain-free state, brownmillerite (BM) $SCO_{2.5}$ is a positive charge transfer antiferromagnetic-insulator[37-41]. However, in the strained state, it turns to a negative charge transfer ferromagnetic-insulator, wherein the O-2*p* hole-mediated exchange between the charge-disproportionated $Co^{3-\delta}$ and $Co^{3+\delta}$ states cause a ferromagnetic ordering in the system[37-41]. The perovskite (P) $SrCoO_3$ ($SCO_3$) on the other hand is a negative charge transfer ferromagnetic-metallic system, which remains stable in the strained state, otherwise, the relaxed $SCO_3$ phase adapts to the $SCO_{2.5}$ phase over the course of time[33]. Thus, from the physical aspects, $SCO_{2.5}$ and $SCO_3$ reveal exclusive structural, electronic, and magnetic properties. The advantage of the SRO layer is its itinerant ferromagnetic character enabling its utility for spintronic applications as well as its usage as a conducting layer. These unique features of $SCO_x$ and SRO provide an attractive combination to design novel spintronic devices as well as fertile land for probing some fundamental research interests.

The reflection high energy electron diffraction (RHEED) intensity oscillations and patterns were monitored in-situ during the growth of the heterostructure as shown in Fig. 1(b). The circular RHEED spots of the Nb-doped STO substrate, recorded just prior to the deposition, become stripes after the heterostructure deposition. Such stripe patterns and intensity oscillations indicate a layer-by-layer growth mode and epitaxial nature of the heterostructure. Here, each oscillation in $SCO_{2.5}$ corresponds to two unit cells (u.c.) due to the alternating stacking of octahedral ($O_h$) & tetrahedral ($T_d$) units along the *c*-direction[37,38,42]. Therefore, 80 oscillations correspond to 160u.c. for the $SCO_{2.5}$ layer. On the top of the $SCO_{2.5}$ layer, we have deposited 9u.c. of SRO.

The $SCO_x$ system can be induced to undergo reversible TPT between $SCO_{2.5}$ and $SCO_3$ phases by the application of electric field[28-30]. However, no such change has been reported for SRO via electric field application. When a potential difference is applied across the interface, with the configuration as shown in Fig. 1(a), though the structure and phase of the SRO layer will remain the same, BM-$SCO_{2.5}$ may undergo a phase transition to P-$SCO_{3-\delta}$. In this context, the metallic SRO acts as a top-electrode layer also. Figure 2(a) shows the room temperature current (I)-voltage (V) relationship of SRO/$SCO_{2.5}$/Nb:STO heterostructure. The I-V behavior reveals an abrupt jump close to 2.1V, in both the increasing and decreasing bias voltage, which corresponds to the TPT between the BM-$SCO_{2.5}$ and P-$SCO_{3-\delta}$[28,29].

The x-ray diffraction (XRD) pattern of the as-grown SRO/$SCO_{2.5}$/Nb:STO heterostructure reveals a single phase nature of the $SCO_{2.5}$ layer in its pristine condition (Fig. 2(b))[28-30,37-39]. The as-grown $SCO_{2.5}$ layer reveals 2-fold superstructure peaks such as (002), (004), (006), (008), and (00<u>10</u>); {(004) and (008) reflections are submerged in the (001) and (002) reflections of the Nb:STO substrate respectively), which arise due to the orthorhombic BM structure having the oxygen defect states[28,29,44] (indexing in Fig. 2(b)). The calculated *c* lattice parameter of the $SCO_{2.5}$ layer is 3.919Å, which is the same as its bulk lattice parameter[39], indicating strain strain-free nature of the $SCO_{2.5}$ layer. On the other hand, the thickness of the SRO layer is only 9u.c. and thus, we could not find any reflection of the SRO layer. However, the obtained ferromagnetic to paramagnetic Curie-temperature ($T_C$ =160K) (as shown later) is very close to the stoichiometric SRO[43], indicating phase pure and stoichiometric growth of SRO in the present heterostructure system. When the potential difference of 2.2V is applied across the interface, which is just above the biasing voltage for inducing the TPT, we notice that the BM phase peaks disappear and new Bragg reflections of (001) and (002) corresponding to the P-$SCO_3$ emerge, confirming the TPT from $SCO_{2.5}$ to $SCO_{3-\delta}$ upon electric biasing[28,29,44]. The calculated *c* lattice parameter 3.829Å of the transformed $SCO_{3-\delta}$ phase is very close to the bulk lattice parameter of stoichiometric $SCO_3$[44], indicating the transformation of the $SCO_{2.5}$ layer to the very near stoichiometric $SCO_3$ layer upon electric biasing. Furthermore, the phase transition is reversible; when the bias voltage is removed, the structure of the $SCO_3$ layer reverts to the $SCO_{2.5}$ phase, as also depicted in Fig. 2(b) (0V data). Moreover, the calculated *c* lattice parameter of this transformed state, after bias removal, is very similar to the as-grown pristine state.

The magnetization of the SRO/$SCO_{2.5}$/Nb:STO heterostructure was investigated across the TPT by recording magnetic moment (M)-temperature (T) behavior in the filed cooled warming (FCW) cycles, as shown in Fig. 2(c). In the absence of the electric bias (0V data, Fig. 2(c)), in the low temperature regime, the temperature dependent magnetization reveals contribution due to the FM nature of the SRO layer, which gradually decreases with increase in temperature. The magnetization reveals upturn above 250K, associated with the antiferromagnetic (AFM) nature of the $SCO_{2.5}$ layer, whose reported Néel-temperature ($T_N$) is 570K[45,46]. Upon the application of 2.2V bias voltage, while we observe a clear ferromagnetic transition of the SRO layer at 160K, which is the same as its bulk counterpart[43], interestingly, it is also noticed that the AFM upturn of the $SCO_{2.5}$ layer changed into FM character with $T_C$ ~320K[47] (2.2V data, Fig. 2(c)). The observed $T_C$ is found to be the maximum among the obtained values in the literature for the $SCO_x$-based materials, indicating transformation to the near stoichiometric $SCO_3$ composition of the $SCO_x$ layer after electric biasing, which is consistent with the XRD data. The maximum $T_C$ obtained in $SCO_{3-\delta}$ single crystal has been 305K[47]. It should be also noted that we could not find any clear $T_C$ corresponding to the SRO layer in the 0V state; possibly due to the overlapping region of FM downturn of the SRO layer and AFM upturn of $SCO_{2.5}$ layer with increasing temperature, which is different from 2.2V bias application condition where both SRO and $SCO_3$ layer depicted FM downturn with increasing temperature.

The observed change in the M-T behavior upon application of such a small potential difference of 2.2 V across the interface leads to the colossal VM effect, which is precisely tunable as well as completely reversible. A completely different M-T behavior of the heterostructure shows a significant magnetic contribution even at room temperature due to the transformation of the AFM $SCO_{2.5}$ to the room temperature FM $SCO_3$. This provides tremendous scope for applications in energy-efficient spintronics. When a potential difference is applied across the heterointerface, the oxygen ions ($O^{2-}$) are interjected from the oxide electrodes (SRO or Nb:STO), depending upon the biasing polarity, into the $SCO_{2.5}$ layer gradually. Accordingly, $SCO_{2.5}$ composition moves towards the $SCO_{3-\delta}$ composition before settling to the oxygen-rich stoichiometric $SCO_3$. This allows us to control the magnetic state of the heterostructure by tuning $\delta$ in $SCO_{3-\delta}$ via potential differences across the heterointerface.

To further demonstrate the colossal VM effect, we recorded the magnetization at -1T applied magnetic field by varying the bias voltage from -3V to +2.7V in the steps of 0.3V at room temperature, as depicted in the left-panel of Fig. 3(a). The measurement protocol is such that we recorded M with elapsed time by increasing applied voltage from -3V to +2.7V in 0.3V steps in every 30s intervals. It is important to note that the M remains stable when the voltage remains the same and changes discretely with changing voltage. We also would like to mention here that the bias response to the M is instantaneous, which is highly appealing for switching device applications. The observed behavior is quite interesting for the VM effect and makes such heterostructures suitable for energy-efficient controllable spintronic applications. The different states of the heterostructure layers and the $O^{2-}$ ion migration in the $SCO_x$ layer in the different biasing conditions are schematically shown in the right-panel of Fig. 3(a)[30].

To further comprehend the VM effect, we plotted the VM-voltage curve in Fig. 3(b), by using the data from Fig. 3(a). We define VM% as VM% $=(M_V-M_0)\times100/M_0$, where $M_V$ and $M_0$ are the moments at the respective applied voltages and 0 applied voltage respectively. By increasing the bias voltage in the positive direction, the VM% increases to about -100% at +2.1V and then it remains almost saturated up to the applied maximum biasing of +2.7V. On the other hand, by increasing the bias voltage in the -Ve direction, the VM increases to about +250% at -2.7V and then it remains almost saturated up to the applied maximum biasing of -3V. It is worthwhile to mention that the different VM% in opposite bias polarity is associated with the different conductivity of the top or the bottom electrodes, which release or receive $O^{2-}$ ions, depending on the bias polarity. This is a crucial finding, since the obtained VM% in the SRO/$SCO_x$/Nb:STO heterostructure is exceptionally high and much simpler to represent than the earlier reported ME coupling in any material. Importantly, the effect is completely robust as can be seen from the recorded M as a function of time with switching voltage between +2.7V and -3V (Fig. 3(c)). The measurement protocol for Fig. 3(c) is also the same as for Fig. 3(a), we recorded M at -1T magnetic field with elapsed time by switching applied voltage between -3V and +2.7V in every 30s interval.

We also made a TPT between $SCO_{2.5}$ and $SCO_3$ in a $La_{0.7}Sr_{0.3}MnO_3$/$SCO_x$/Nb:STO heterostructure and in this system also we observed a similar colossal VM effect (not shown here). Moreover, because of the reversible TPT nature between the $SCO_{2.5}$ and $SCO_3$ phases, the physical states of the metallic-magnet/TPT-material heterostructure can be set and reset by applying an appropriate potential difference across the heterointerface. Therefore, this work is not only rich from a fundamental physics point of view but also beneficial for a broad region of energy-efficient and scalable spintronic applications. This work also leads to encouraging exploring other combinations of metallic-magnet/TPT-material heterostructure systems[48-53] for obtaining better VM%.

In conclusion, we showed that utilizing TPT in a metallic-magnet/TPT-material heterostructure through an applied electric field is an alternative knob to tune the magnetic properties in all oxide heterostructures, which has been so far scarcely explored. A tremendous change in the electronic and magnetic properties occur across the TPT in such a heterostructure because of the tuneable interfacial oxygen stoichiometry reversal with the application of a very small electric field. It shows that the observed volto-magnetic effect is robust and significant compared to the magneto-electric effects observed in composite materials arising due to the charge accumulation or depletion at the interface, strain-mediated effects, ion migration, and morphological changes in any material. Our results encourage exploring element-resolved spectroscopic/microscopic investigation in such metallic-magnet/TPT-material heterostructures for better optimizing VM%, making an all-oxide metallic-magnet/TPT-material heterostructure system a promising candidate for the future low-energy consuming spintronic application. Moreover, electronic and magnetic structure probes with sufficient spatial resolution will enhance the functionality of such heterostructure systems to integrate them into miniaturized devices.

## Methods

**Sample growth.** Epitaxial heterostructure of SRO/$SCO_{2.5}$ was grown on conducting Nb-doped STO (001) substrate by pulsed laser deposition (PLD) technique. The PLD system is a KrF excimer laser (Coherent, USA), $\lambda$ =248nm, pulse width 20ns, assisted by the in-situ RHEED (Staib Instruments, USA). The well-sintered dense pallets of $SCO_{2.5}$ and SRO prepared using solid state reaction route were used as the targets for the respective thin film depositions. During the deposition, substrate temperature was kept at 750°C, oxygen partial pressure was maintained at 200mTorr, laser energy density at the target was 1.8J/cm$^2$ for $SCO_{2.5}$ as well as SRO, and target to substrate distance was kept at 6cm. After deposition, the grown heterostructure was cooled in 200mTorr of oxygen partial pressure.

**I-V measurements.** A homemade resistivity set-up was used to record the I-V behaviour of the heterostructure.

**Structural characterizations.** Bruker D2-Phaser x-ray diffractometer (Cu $K_\alpha$: $\lambda$ =1.5406 Å) was employed for structural characterization of the grown heterostructure without and with electric field using a home-built sample station enabling electric field dependent XRD measurements.

**Magnetic measurements.** We also used a home-made sample insert for performing in-situ electric field-dependent magnetic measurements using a 7T SQUID-VSM system (Quantum Design, USA).

### Author contributions

S.C. and R.J.C. conceived the concept and course of the problem and designed the experiments. S.C. performed the sample preparation and RHEED measurements with assistance from R.M. and A.B. S.C. acquired the XRD data. S.C., A.K.M., S.Y., and S.K. performed the I-V measurements. S.C., S.M., A.B., and R.M. collected and analyzed the magnetization data with assistance from D.M.P. and R.J.C. S.C. wrote the manuscript with R.J.C., using substantive feedback from all others.

### Acknowledgements

The authors are thankful to DST-SERB for the grant under the project CRG/2021/001021. We also acknowledge Moritz Hoesch for the valuable discussion.

**Figure captions:**

Figure 1. (a) Schematic of metallic-magnet/TPT-material/Nb:STO heterostructure device. (b) RHEED intensity oscillations during the heterostructure deposition. In the inset, the images on the left and the right are the RHEED patterns of the Nb:STO substrate, recorded just prior to the deposition, and just after the heterostructure deposition, respectively.

Figure 2. (a) Room-temperature I-V characteristics of SRO/$SCO_x$/Nb:STO heterostructure in both increasing and decreasing voltages. The jump in current at ~2.1V arises because of the TPT. The inset shows the schematic crystal structures of the $SCO_x$ layers across the TPT. The green and the blue blocks are the $O_h$ and $T_d$ units respectively. BM-$SCO_{2.5}$ structure possesses alternate repetition $O_h$ and $T_d$ units along the *z*-direction because of the 1-*d* oxygen vacancy channel ordering. Above the TPT voltage 2.1 V, the $T_d$ units become $O_h$ units as a result of $O^{2-}$ intercalation and leave only $O_h$ units in the purely stoichiometric P-$SCO_3$ phase. (b) Bias-dependent XRD of the SRO/$SCO_x$/Nb:STO heterostructure. The peak corresponds to brownmillerite and perovskite phases are designated as BM and P respectively. (c) Bias-dependent FCW M-T behavior recorded at 1000 Oe magnetic field of the SRO/$SCO_x$/Nb:STO heterostructure.

Figure 3. (a) (Left-panel) Room temperature bias dependent M behavior of the SRO/$SCO_x$/Nb:STO heterostructure device. (Right-panel) schematic view of $O^{2-}$ ion migration in the $SCO_x$ layer, depending on the bias. (b) VM%-voltage behavior extracted from Fig. 3(a) data. (c) Room temperature magnetic switching behaviour with the bias in the SRO/$SCO_x$/Nb:STO heterostructure device.

**Figures:**

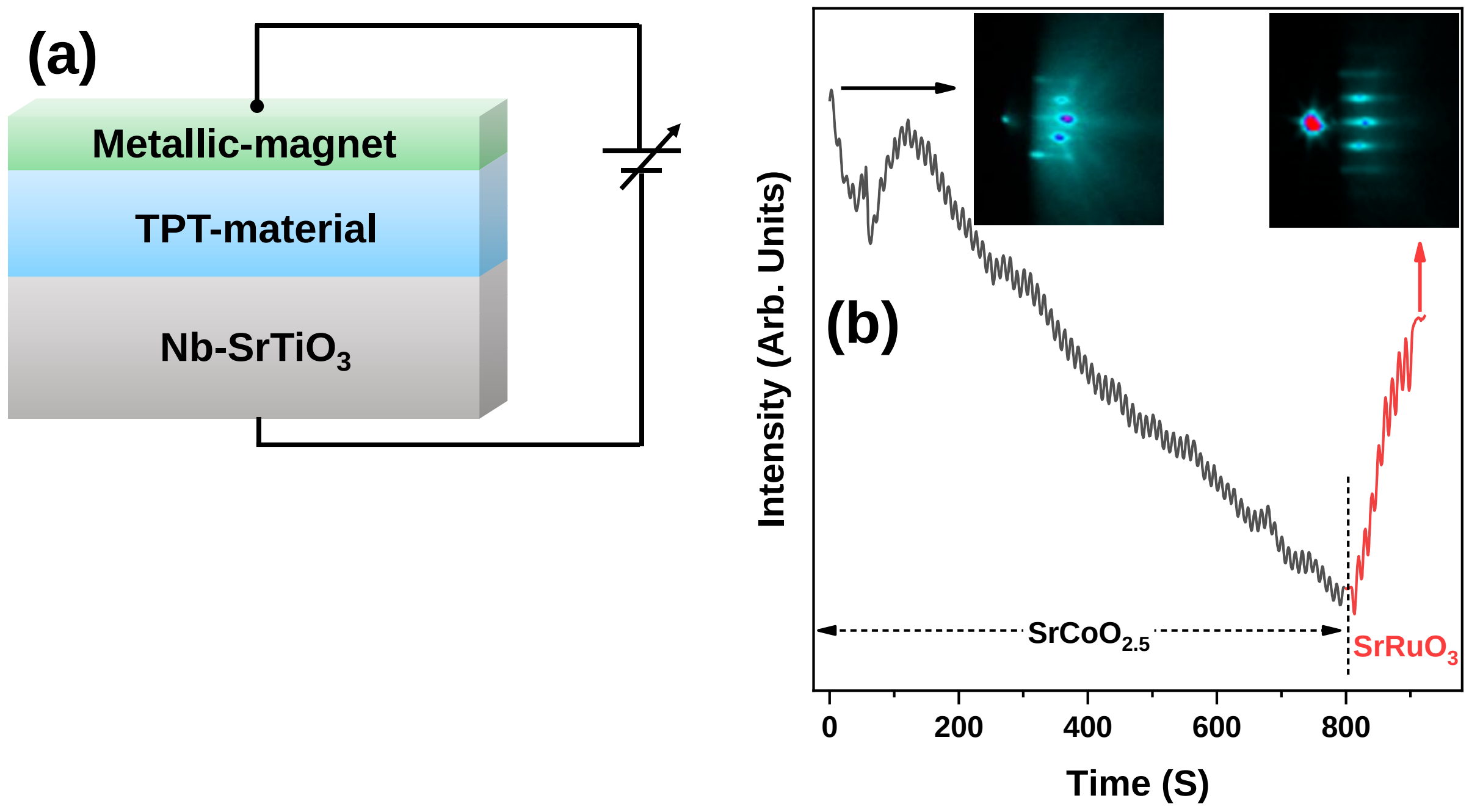

**Fig. 1**

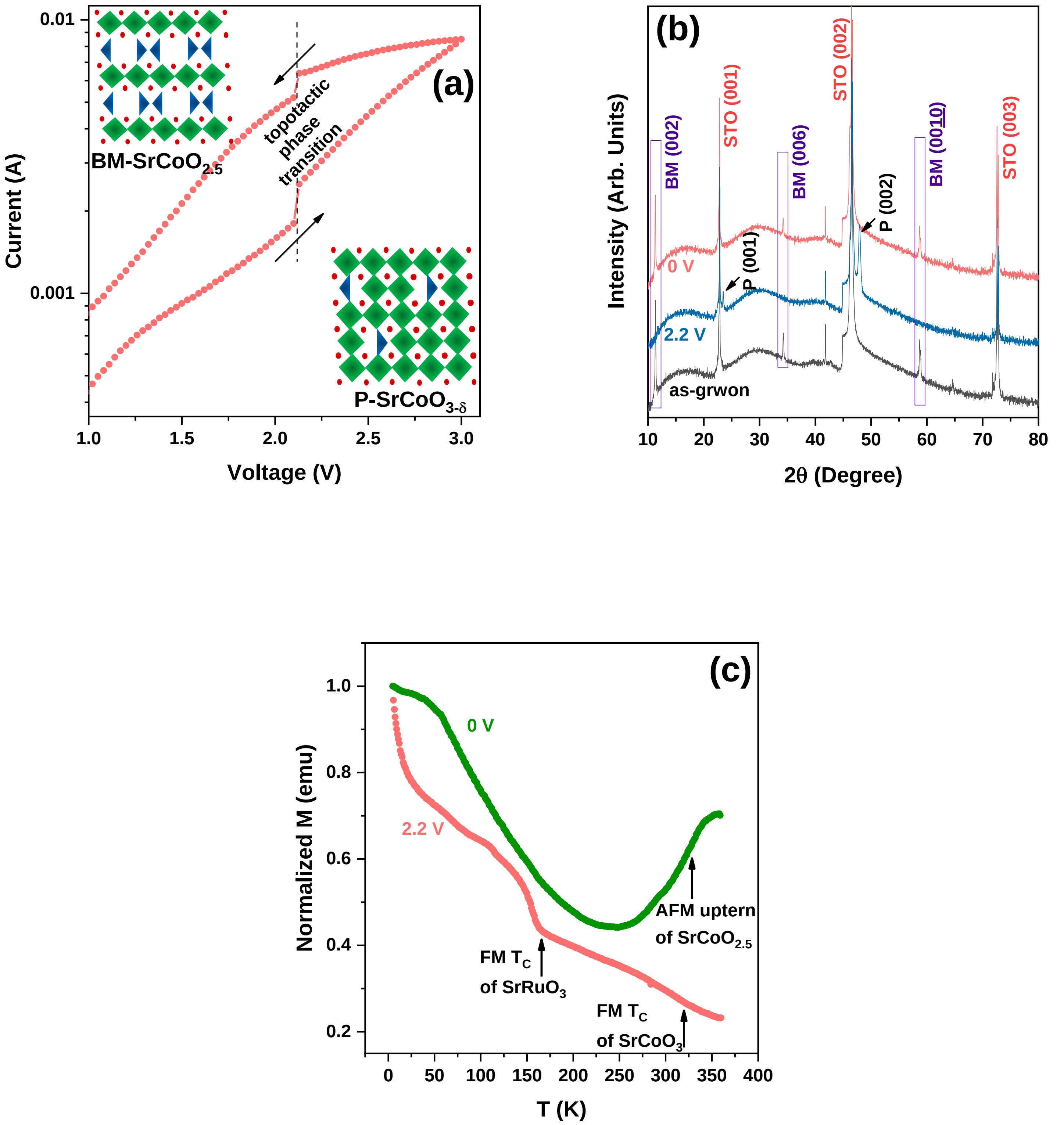

**Fig. 2**

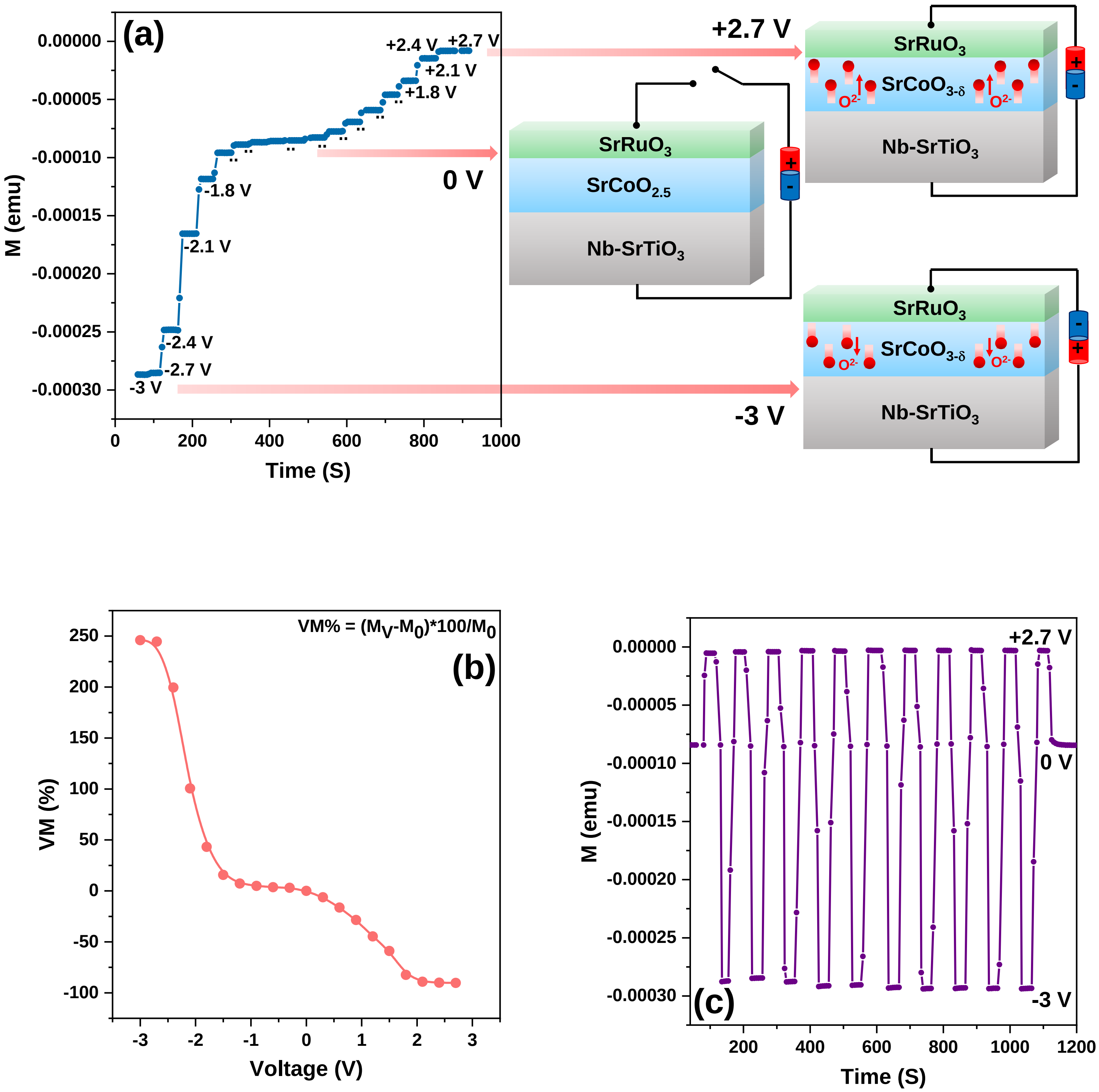

# Fig. 3